\documentclass[]{spie}  %>>> use for US letter paper
\usepackage{amsmath,amsfonts,amssymb}
\usepackage{graphicx}
\usepackage[colorlinks=true, allcolors=blue]{hyperref}

\title{Atmospheric dispersion correction: \\ Model requirements and impact on radial velocity}

\author[a,b]{Wehbe, B.}
\author[c,d]{Cabral, A.}
\author[e,a]{Figueira, P.}
\author[f]{\'{A}vila, G.}
\affil[a]{Instituto de Astrof\'{i}sica e Ci\^{e}ncias do Espa\c{c}o, Universidade do Porto, CAUP, Rua das Estrelas, 4150-762 Porto, Portugal}
\affil[b]{Departamento de F\'{i}sica e Astronomia, Faculdade de Ci\^{e}ncias, Universidade do Porto, Rua Campo Alegre, 4169-007 Porto, Portugal}
\affil[c]{Instituto de Astrof\'{i}sica e Ci\^{e}ncias do Espa\c{c}o, Universidade de Lisboa, Campus do Lumiar, Estrada do Pa\c{c}o do Lumiar 22, Edif. D, PT1649-038 Lisboa, Portugal}
\affil[d]{Departamento de F\'{i}sica, Faculdade de Ci\^{e}ncias, Universidade de Lisboa, Campo Grande 1749-016 Lisboa Portugal}
\affil[e]{European Southern Observatory, Alonso de C\'{o}rdova 3107, Vitacura, Santiago, Chile}
\affil[f]{European Southern Observatory, Karl-Schwarzschild-Stra{\ss}e 2, 85748 Garching bei M\"{u}nchen, Germany}

\authorinfo{E-mail: bachar.wehbe@astro.up.pt}

\begin{document} 
\maketitle

\begin{abstract}
Observations with ground-based telescopes are affected by differential atmospheric dispersion when seen at a zenith angle different from zero, a consequence of the wavelength-dependent index of refraction of the atmosphere. One of the pioneering technology in detecting exoplanets is the technique of radial velocity (RV), that can be affected by uncorrected atmospheric dispersion. The current highest precision spectrographs are expected to deliver a precision of 10 cm s$^{-1}$ (e.g., ESPRESSO). To minimize the atmospheric dispersion effect, an Atmospheric Dispersion Corrector (ADC) can be employed. ADC designs are based on sky dispersion models that nonetheless give different results; these can reach a few tens of milli-arcseconds (mas) in the sky (a difference up to 40 mas); a value close to the current requirements (20 mas in the case of ESPRESSO). In this paper we describe tests done with ESPRESSO and HARPS to understand the influence of atmospheric dispersion and its correction on RV precision. We also present a comparison of different sky models, using EFOSC2 data (between 600nm and 700nm), that will be used to improve on the design of ADCs. 
\end{abstract}


\section{INTRODUCTION}
\label{sec:intro}  % \label{} allows reference to this section

\begin{figure}[ht]
\centering
\includegraphics[scale=0.2]{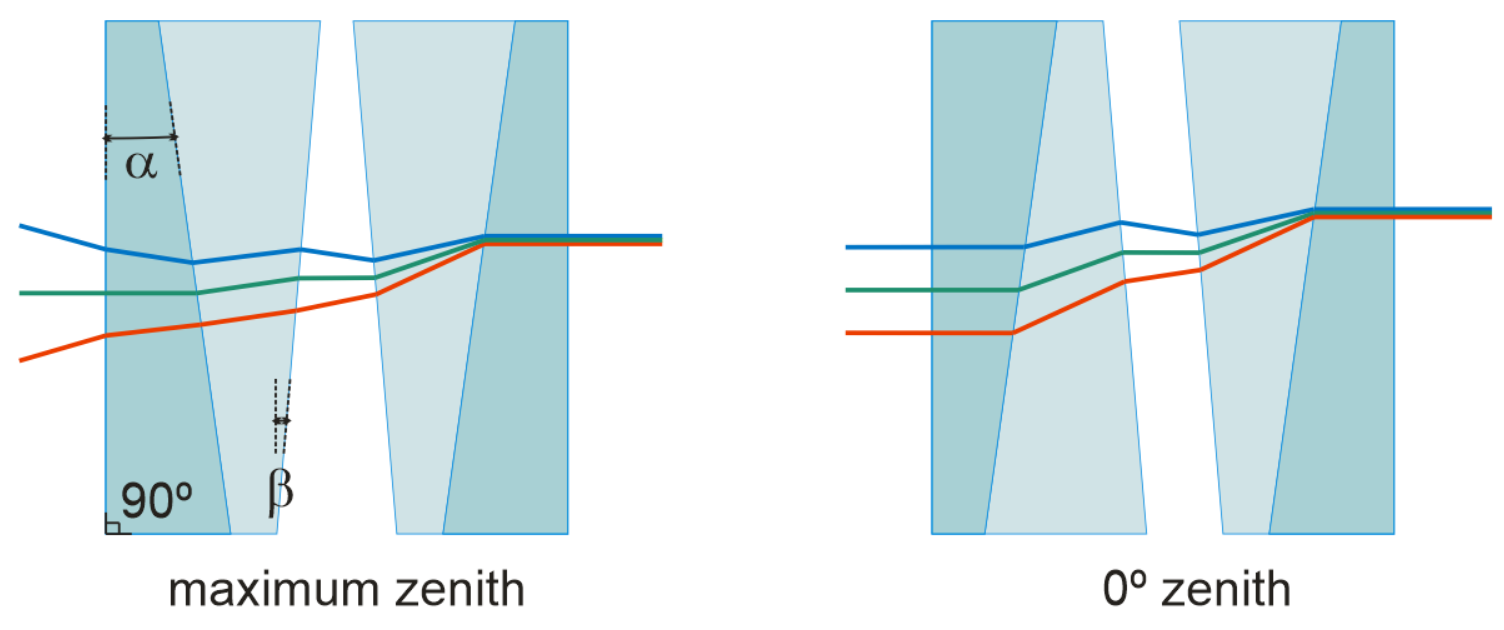}
%\resizebox{\hsize}{!}{\includegraphics{adc.png}}
\caption{ADC with counter rotation prisms, for a maximum zenith (maximum zenithal angle) (left) and 0$^{\circ}$ zenith (right) configuration \cite{Cabral2012}. $\alpha$ and $\beta$ are the 2 prism angles. The blue, green and red lines represent different wavelengths.}
\label{fig:adc}
\end{figure}

\begin{figure}[ht]
\centering
\includegraphics[scale=0.25]{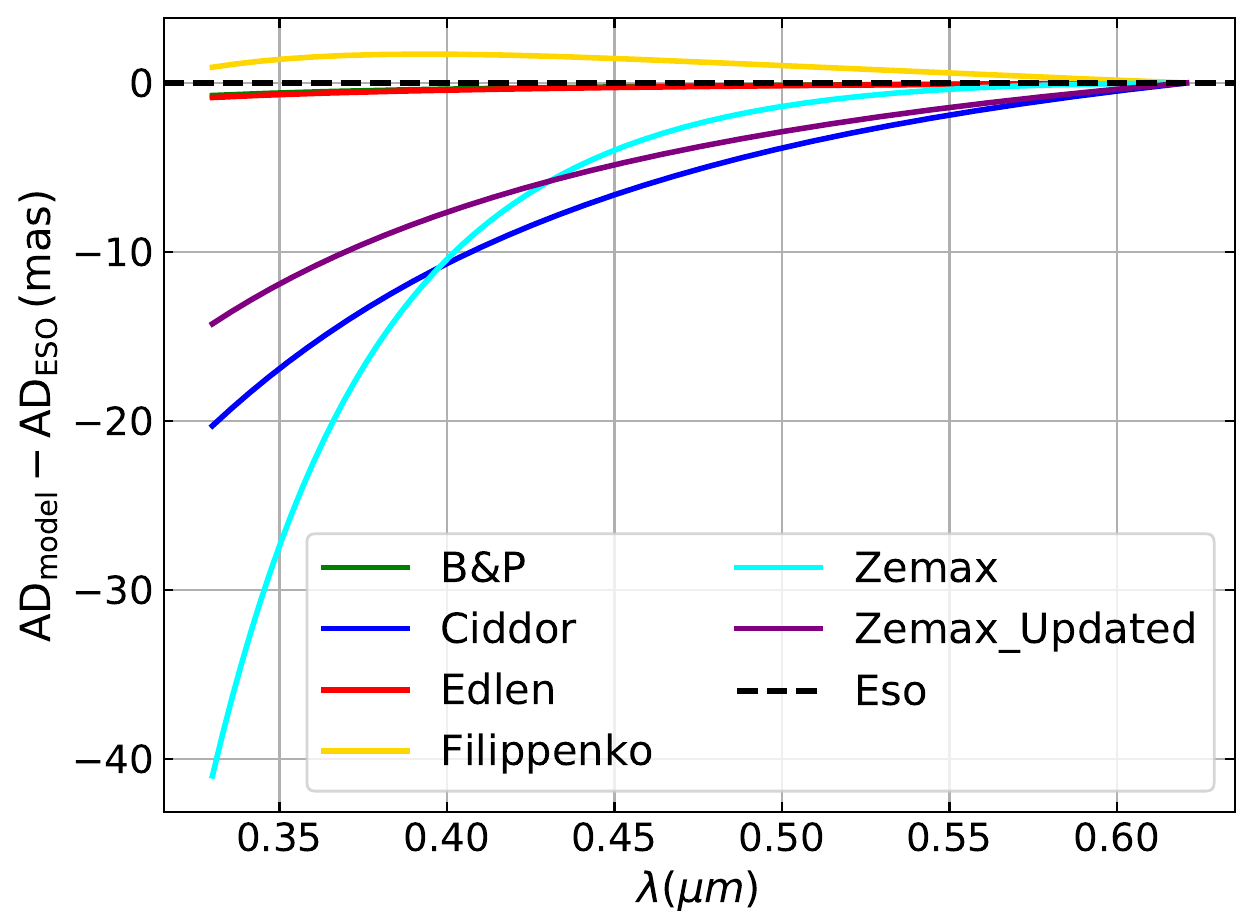}
%\resizebox{\hsize}{!}{\includegraphics{models.png}}
\caption{Differential atmospheric refraction for a zenith angle of 60$^{\circ}$ at Paranal for different models using same atmospheric conditions. The ESO model was used as reference.}
\label{fig:models}
\end{figure}

Astronomical observations performed using ground-based telescopes are affected by wavelength-dependent atmospheric dispersion when seen at a zenith angle different from zero. The atmospheric dispersion is due to the variation of the refractive index of the atmosphere. This variation has been discussed by several authors: B\"{o}nsch \& Potulski \cite{Bonsch1998}, Ciddor \cite{Ciddor1996}, Edl\'{e}n \cite{Edlen1966}, Filippenko \cite{Filippenko1982}. By definition, an atmospheric dispersion model computes the refractive index of dry air at sea level in function of wavelength, n($\lambda$). Since observatories are usually located at higher altitudes, the refractive index should be corrected for temperature, pressure and relative humidity (RH). Differential atmospheric dispersion is then calculated using equation \ref{eq:dispersion} \cite{Smart1931}

\begin{align}
\Delta R(\lambda) &= R(\lambda) - R(\lambda_{\rm ref}) \nonumber  \\
\Delta R(\lambda)  &\approx 206265 \left[ n(\lambda) - n(\lambda_{ \rm ref}) \right] \times \tan z,
\label{eq:dispersion}
\end{align} 
where  $\lambda_{\rm ref}$ is the reference wavelength, and $z$ is the zenith angle of observation. 
A good review of these models and their difference is provided by Span\`{o} \cite{Spano2014} who compared the air refractivity of several models with the Zemax model \cite{Zemax} and provided an updated version, hereafter named ``Zemax updated". To our knowledge, these atmospheric models have never been directly compared to on-sky measurements. Nonetheless, these models have seen wide usage on the development of state-of-the-art instrumentation, and have been used to place high constraints on the technical requirements, like in ESPRESSO \cite{Megevand2014}. \\
Even assuming a perfect telescope centering and guiding at a given reference wavelength, atmospheric dispersion will shift the image centroid as a function of wavelength. This effect has a very strong impact on the light collected by a spectrograph, especially in high-resolution spectrographs aiming at high fidelity spectra. The loss of light as a function of wavelength introduces a wavelength dependent photon-noise that becomes particularly stronger as we go towards the blue wavelengths where the dispersion is stronger, reducing our ability to measure efficiently the line's position and its properties. In addition, imperfect atmospheric dispersion will introduce a varying slope in the spectral continuum \cite{Pepe2008} which will alter the weight of spectral lines and influence the mean radial velocity (RV) value.
\\
This problem is particularly serious for high-resolution spectrographs aiming at sub-pixel RV precision. With the objective of developing high-resolution spectrographs that can reach an RV precision \cite{Fischer2016} of \newline 10 cm s$^{-1}$, several instrumentation challenges need to be solved including a correction of atmospheric dispersion variation down to this value. In order to reach the highest reproductibility in collected spectra and RV properties, an Atmospheric Dispersion Corrector (ADC) is thus mandatory.
\\
The ADC is the first optical element of the spectrograph's front end, meaning that the spectrograph will disperse the image of the star as seen through the ADC. The correction of atmospheric dispersion will correct the lambda-dependent position of the star on the sky. In turn, this will lead to a correction of the flux variation induced by absorption. The typical configuration of an ADC is two counter-rotating prisms (Fig.\ref{fig:adc}) able to compensate the atmospheric dispersion by reproducing the opposite effect to that of the atmosphere.
\\
The design and operation of ADCs is based on the aforementioned atmospheric models, that lead to different results for the same input values (wavelength, pressure, temperature, RH, zenith angle) as seen in Figure \ref{fig:models}. The difference between the various models is particularly severe in the blue part of the spectrum where the atmospheric dispersion is larger. For example, for a zenith angle of 60$^{\circ}$, the difference between some of the most used models (Zemax and the ESO model, for example) is as large as 45 milli-arcseconds (mas).
\\
The requirements on ADC operation cannot be smaller than the accuracy of the ADC at reproducing the real dispersion. Imposing more demands on the ADC's operations and performances is becoming a real challenge in terms of atmospheric dispersion.
\\ Thus it is fundamental to measure the atmospheric dispersion on-sky to assess the accuracy of these models and be able to understand which model better mimics the sky, in order to develop better ADCs to reduce the errors on RV, and thus enable further discoveries of potentially habitable worlds.

\section{Impact of atmospheric dispersion}
\subsection{Using trace to measure dispersion}
Skemer, A.J. et al. (2009) \cite{Skemer2009} showed that atmospheric dispersion is an important limitation to image quality for the extremely large telescopes (primary mirror of 20 m up to 100 m) since it will severely affect the Strehl ratio (a measure of the quality of optical image formation) and the full width at half maximum (FWHM) of the images. They tried to measure the atmospheric dispersion in the N-band, where the atmospheric dispersion is as low as 5 mas, and compared their results to atmospheric models. Their idea was to use the order position measured on the detector as direct measurements of the atmospheric dispersion. Spectroscopic measurements have the advantage of measuring all the wavelengths of interest simultaneously so that the overall trend and curvature of the atmospheric effect through the range of interest is unambiguous.
\\

Without an ADC, the image of a target, dispersed through the atmosphere, is elongated. At an airmass of 1 ($z$ = 0$^{\circ}$), the atmospheric effect is null. At this position, the image of a target should be a point source, only dispersed by the grism of the spectrograph. A high resolution spectrograph is composed of two dispersive elements; the main and most efficient dispersive element is the diffraction grating creating the main dispersion and a secondary less powerful dispersive element to cross-disperse the orders which is called the secondary dispersion \cite{Figueira2018}. The use of a slit spectrograph, with the slit directed along the dispersion direction (the parallactic angle direction), will result in a spectrum with a dispersion induced by the spectrograph's main dispersive element, creating the spectrum along the wavelength direction, and a dispersion along the spatial direction, created by the atmospheric dispersion. The light is dispersed in a unique direction. If any element creates an additional dispersion, we can use observations at different airmasses to statistically determine the dispersion and any instrumental offset, between the observations and the models.
\\
In order to measure any effect the instrument might create, we fit the trace of the spectrum at an airmass of 1, repeat the trace measurements for targets at different airmass, and subtract any instrumental effect. This will allow us to directly measure the atmospheric dispersion across the spectral range and compare it to the various proposed atmospheric models, using the same atmospheric parameters as the ones of the observation time.

\subsection{Impact on the radial velocity}
\begin{figure}[ht]
\centering
\includegraphics[scale=0.4]{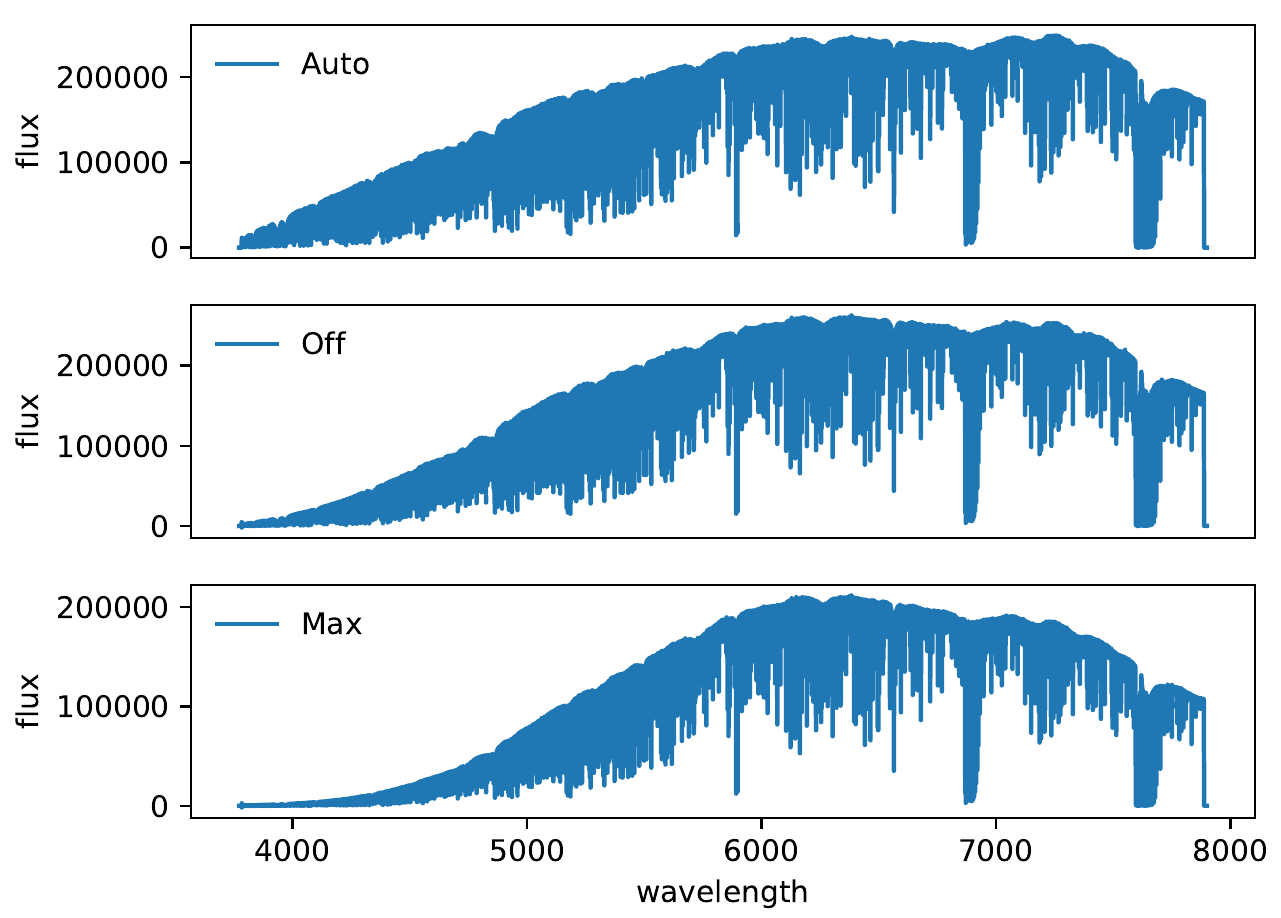}
%\resizebox{\hsize}{!}{\includegraphics{slope_variation_ut2_hd1581.png}}
\caption{Slope variation in function of dispersion (different positions of the ADC). The variation is more severe on the blue part of the spectrum due to bigger dispersion. Top: The ADC is at Auto mode (dispersion ~0); middle: the ADC is Off (dispersion is only due to the atmosphere: 1.14"); bottom: the ADC is at Maximum (dispersion is double that of the atmosphere: 2.24")}
\label{fig:slope}
\end{figure}

Pepe, F. et al. \cite{Pepe2008} presented several instrumental effects that might contaminate the target spectrum and alter the computed RV. An imperfect atmospheric dispersion correction will introduce a slope variation in the spectral continuum (see Fig. \ref{fig:slope}). This variation, will lead to a different weight for the spectral lines which will impact on the final computation of the RV. In order to prevent and correct this effect, the authors suggested the following:
\begin{itemize}
    \item the use of an ADC,
    \item correct the slope variation in the data reduction pipelines.
\end{itemize}
An ADC is currently being used in most high-resolution astronomical instruments to reduce the atmospheric dispersion down to the level of tens of mas. In addition, a flux correction function is being used in some reduction pipelines aiming at a higher-level chromatic correction. Due to dispersion, parts of the spectra might be lost\cite{Pepe2008}\cite{Fischer2016}. This loss is causing slope variation in the continuum. In order to calibrate the spectrum, the detected flux should be multiplied by a wavelength-dependent factor that will set the level of the continuum to the same level as the expected absolute flux above the atmosphere. Using an exposure time calculator, one can compute the expected counts of the target taking into consideration the sky conditions and the instrument setup. Then by measuring the detected flux on the detector, one can correct for any flux losses by multiplying by the ratio of expected over observed (measured) flux::
\begin{align}
R = \frac{F_{\rm exp}}{F_{\rm obs}}
\label{eq:flux_corr}
\end{align} 

\section{Observations and data analysis}
\subsection{Observations and data analysis: Models and trace analysis}
In order to measure the atmospheric dispersion with the method previously described, a slit spectrograph is needed. Since the dispersion is proportional to angular magnification of the telescope \cite{Atad2008}, the bigger the telescope is, the easier it will be to detect the effect of dispersion. Most of the new spectrographs are equipped with an ADC, which will correct the effect of the atmosphere we are aiming at measuring. So in order to fulfill our requirements, a slit spectrograph that isn't equipped with an ADC is required to perform the task. The spectrograph EFOSC2 \cite{Buzzoni1984}, the ESO Faint Object Spectrograph and Camera (v.2), can fit these requirements. In fact, EFOSC2 is a slit spectrograph mounted at the Nasmyth B focus of the NTT.
\\
We browsed ESO's archive in order to find data for different targets at different zenith angles taken through the same night. Using observations from a single night will minimize the effects of variable seeing on the PSF of the star as seen through the spectrograph. We also made sure that the observations were made with the same optical setup in terms of grisms (grism \# 20; 600 nm $\leqslant$ $\lambda$ $\leqslant$ 700 nm) and filters (no filter). Table \ref{table:targets} summarize the targets and the atmospheric conditions at the time of observations.

\begin{table}[ht]

\caption{EFOSC2 targets and weather data (August 17-18, 2011)}
\centering
\begin{tabular}{l c c c c}
\hline
Object & Airmass & T ($^{\circ}$C) & P (mbar) & RH (\%) \\
\hline
HD 172190 & 1.011 & 7.85 & 768.5 & 19 \\
HD 207155 & 1.19 &  7.6 & 768.4 & 20 \\
HD 24587 & 1.237 & 7.15 & 768.8 & 18 \\
HD 29248 & 1.39 & 7.45 & 768.7 & 16 \\
HD 23466 & 1.686 & 6.8 & 769 & 22 \\
HD 26677 & 1.738 & 7.2 & 768.7 & 17 \\
\hline
\end{tabular}
\label{table:targets}
\end{table}

\begin{figure}[ht]
    \centering
    \includegraphics[scale=0.4]{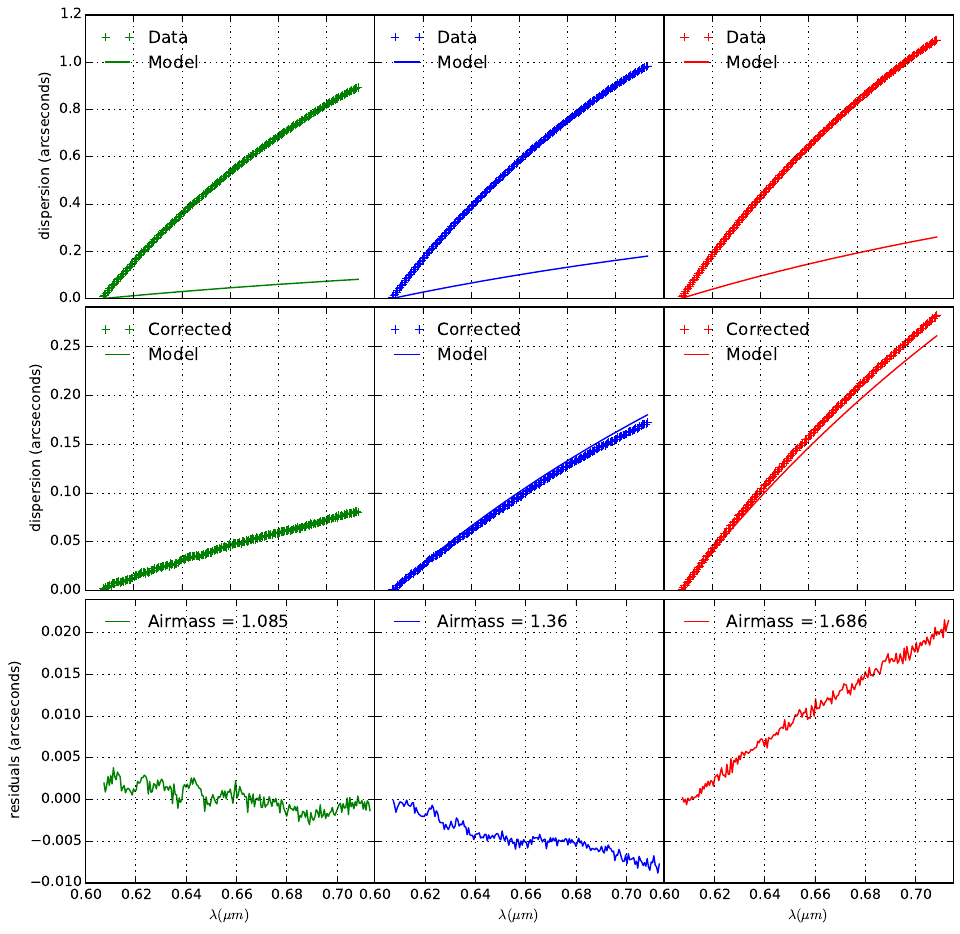}
    %\resizebox{\hsize}{!}{\includegraphics{details_trace.png}}
    \caption{Results of the method tested on targets at different airmass using EFOSC2 data (Top: Atmospheric dispersion \texttt{vs} wavelength (data: measured; model: Filippenko); middle: data corrected from instrumental effect \texttt{vs} wavelength; bottom: residuals of corrected data and model.)}
    \label{fig:details_trace}
\end{figure}

The weather data listed in Table \ref{table:targets} is extracted from the headers of each file. These same data are used when computing the atmospheric dispersion using the models mentioned before.
\\
In order to extract any instrumental effect, we have to trace a target observed at zenith. In our case it was HD 172190 observed at an airmass of 1.011.
\\
From an optical point of view, the instrumental effects are supposed to be independent of the zenithal angle of observation. In that case, for a given wavelength, the instrumental effects should be constant for all targets. The difference between the trace of the targets, and the instrumental effect, should represent the atmospheric dispersion. After estimating the instrumental effects, it can be subtracted from the trace of all the targets. The resulting plot should represent the atmospheric dispersion taking into consideration the atmospheric parameters. In Figure \ref{fig:details_trace}, we show the results of the method tested on targets at different airmass using the EFOSC2 data. Once the trace measurements were corrected for the instrumental effect, it is clear from panel 2 of Figure \ref{fig:details_trace} that the Filippenko's model is able to reproduce the data.

\begin{figure}[ht]
    \centering
    \includegraphics[scale=0.4]{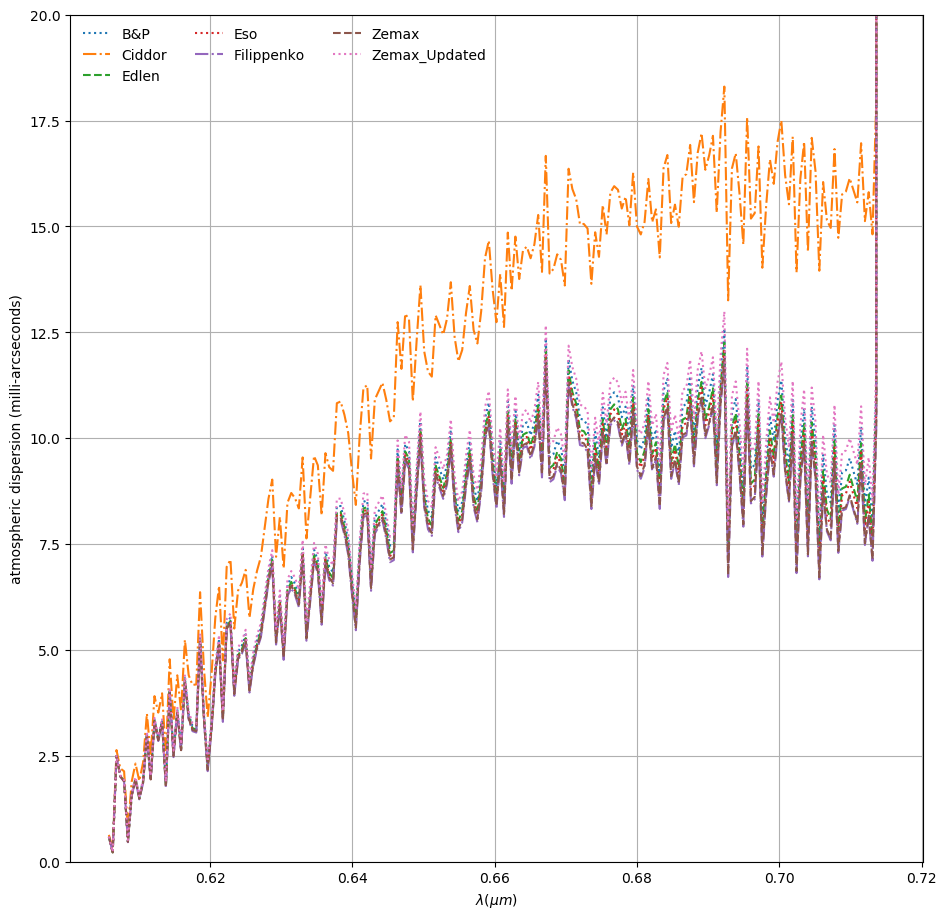}
    %\resizebox{\hsize}{!}{\includegraphics{models_residuals.png}}
    \caption{Residuals of corrected data and all the models. The case of HD 23466 (airmass of 1.686)}
    \label{fig:models_residuals}
\end{figure}

In order to compare the final dispersion extracted with the modeled one, we calculated the residuals between the model and the extracted dispersion (bottom panel of Figure \ref{fig:details_trace}). We found out that the Filippenko's model is able to reproduce the data up to an accuracy of $\pm$ 20 mas. The difference in the residuals clear in Figure \ref{fig:details_trace}, could be interpreted in two different scenarios: it could be that the model is not able to reproduce the exact atmospheric dispersion, or that the residuals are instrumental and not related to the model.
\\
If it was the first case, one would expect that all the residuals should have the same shape. This is not the case. Going further into the details of EFOSC2, we found out that the default slit alignment is in the East-West direction. In order to keep the slit aligned along a given Position Angle, an offset of \newline PA + 90$^{\circ}$ must be applied. This offset is done using a rotator that has an accuracy of 0.1$^{\circ}$. This accuracy introduces a misalignment between the slit and the atmospheric dispersion direction which may be the factor responsible of these residuals. On the other hand, when observing close to the Zenith, the telescope and the rotator have to rotate very fast in order to track the object. The result will be a poor image quality, less accurate pointing and problems with the positioning of the targets on the slit. Having a slit that is not perfectly aligned with the atmospheric dispersion direction will introduce an extra slope while tracing the spectroscopic order. This extra slope, is most probably the reason why the model is sometimes not able to represent perfectly the data and returning residuals up to 20 mas. We also computed the residuals from all the aforementioned atmospheric models (see figure \ref{fig:models_residuals}). At this level of accuracy, we cannot confirm which model is better representing the data. Further investigations are certainly needed.

\subsection{Observations and data analysis: HARPS and ESPRESSO RV tests}
In order to understand the effect of atmospheric dispersion on RV measurements, two tests were done: one using HARPS data from the archive, to test the effect of the flux correction function by turning it on/off, and another one using ESPRESSO data that we got with different position of the ADC.
\\
\subsubsection{HARPS}
\begin{figure}[ht]
    \centering
    \includegraphics[scale=0.4]{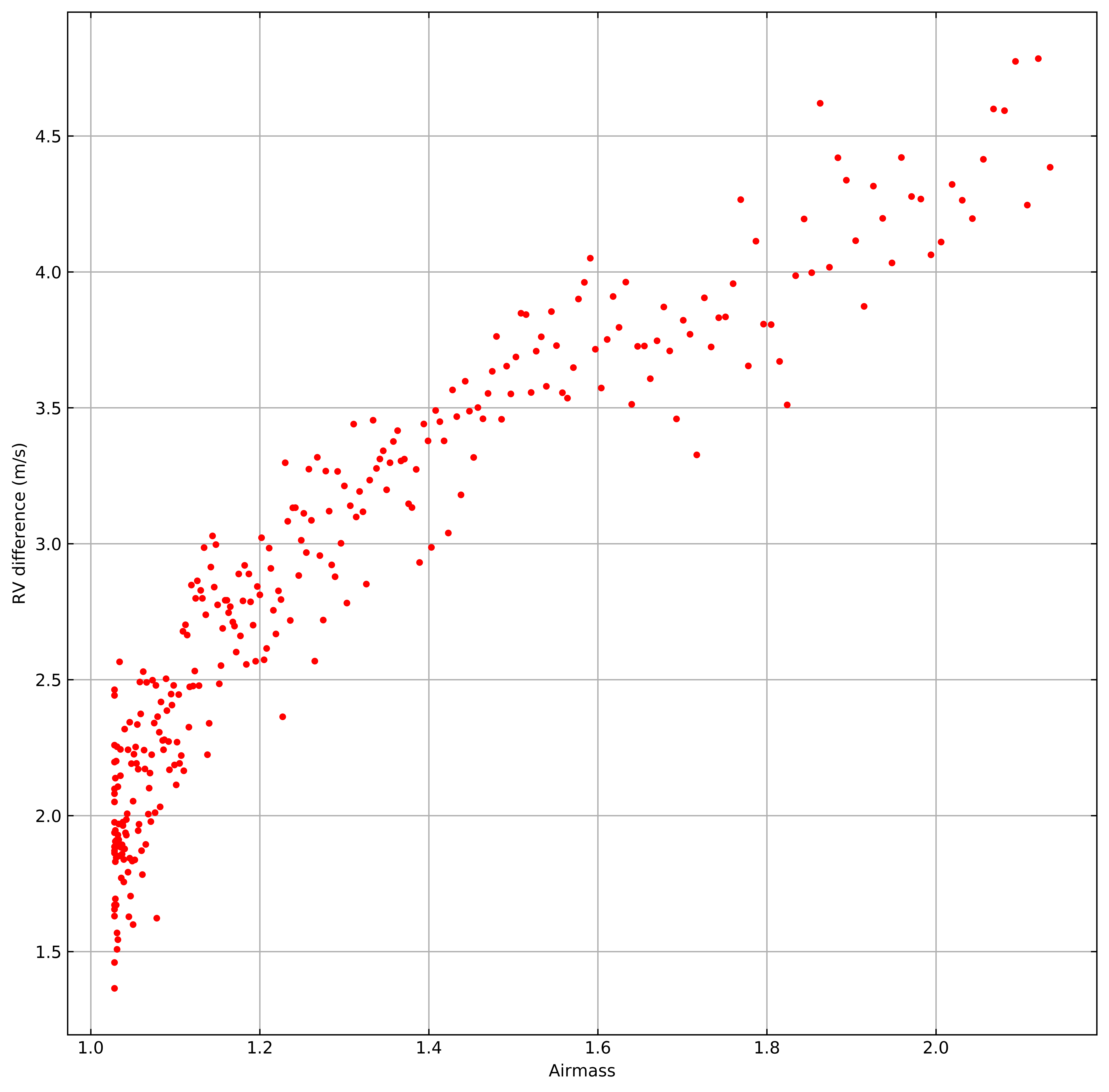}
    %\resizebox{\hsize}{!}{\includegraphics{harps.png}}
    \caption{Difference in computed RV between turning on and off the flux correction in function of airmass. The case of HD 10700.}
    \label{fig:harps}
\end{figure}

In order to evaluate the effect of the flux correction function, we browsed the HARPS ESO archive to find suitable data for our test. HARPS is equipped with an ADC, so what we will be testing using the function is the effect of residuals on RV. We also know that the residuals will increase with airmass. To perform our test we selected targets observed at different airmasses through one night, with the same instrumental setup. HD10700, was observed in the night of 2004-10-05 \cite{Teixeira2009}. The star was followed through the whole night and covered a range of airmass between 1 and 2.2. We reduced the data using the HARPS pipeline (version 3.5) in order to compute the RV. Two sets of reduced data were done by turning on and off the flux correction function, $(\tt{ic\_do\_flux\_correction})$, which is a pipeline variable. The difference between them is then plotted as function of airmass in figure \ref{fig:harps}. This figure shows that, as expected, the difference in RV between flux correction on and off increases with airmass, hence with dispersion. With the increase of airmass, the amount of atmospheric dispersion increases rapidly. The ADC is able to correct up to an airmass of 2. Above that, the amount of residuals will increase which explains the disparity in the RV difference.
\subsubsection{ESPRESSO}
\begin{figure}[ht]
    \centering
    \includegraphics[scale=0.75]{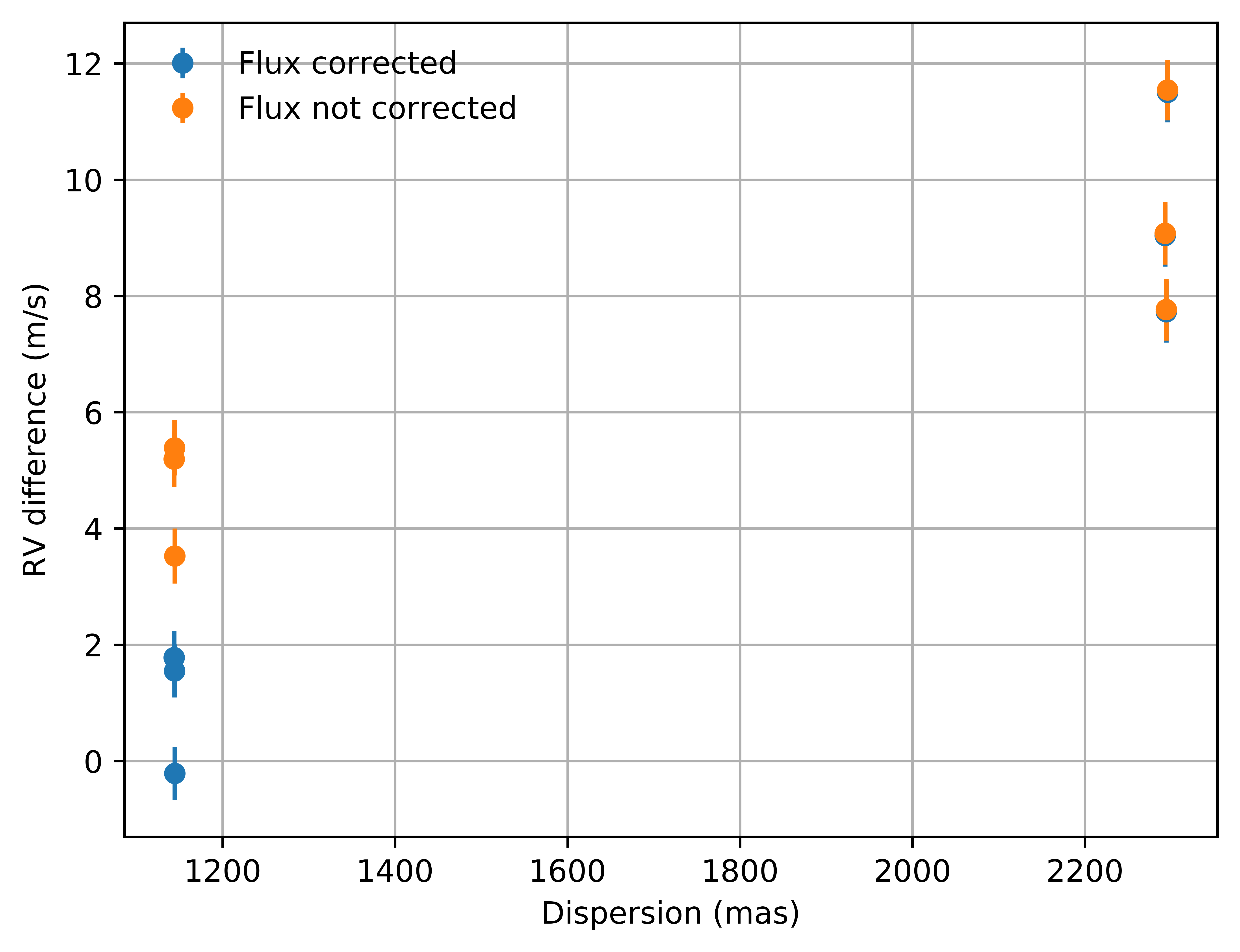}
    %\resizebox{\hsize}{!}{\includegraphics{disp_rv_UT2.png}}
    \caption{Difference in RV between ESPRESSO's ADC Off and Max compared to ADC Auto, when the flux correction function was on and off.}
    \label{fig:disp_rv}
\end{figure}

We performed the same tests on ESPRESSO as we did on HARPS with the difference of changing the positions of the ADC. The target observed was HD 1581 while setting the ADC at different rotational angles:
\begin{enumerate}
    \item Auto: The ADC is set to Auto mode. This means that the ADC will rotate and correct the atmospheric dispersion. The dispersion residuals will be $\sim$ several mas.
    \item Off: The ADC is set to Off mode. In this case the the 2 prisms of the ADC are set with their apex opposite to each other. The ADC will not correct the atmospheric dispersion in this case, neither introduce any extra dispersion. The total dispersion is equal to the atmospheric one.
    \item Max: We forced one of the prisms to be at 180$^{\circ}$ for its Auto position. In that case the ADC will introduce an extra dispersion. The total dispersion is two times that of the atmosphere.
\end{enumerate}
By setting the ADC at different positions, we introduce different levels of dispersion. By turning on and off the slope correction function, we will test the effects of dispersion on the final RV. These 3 cases will also give us an idea of the effects of dispersion on the signal to noise ratio (SNR). Figure \ref{fig:disp_rv} depicts the results.
\\

In the case of ESPRESSO tests, by RV difference we mean the difference in the computed RV between the different cases of the ADC compared to when it is at Auto mode i.e., when the dispersion is at its minimum. As expected, this difference is increasing with dispersion, and is larger when the flux correction function is off. This is clear in figure \ref{fig:disp_rv}. It can also be seen that at the higher dispersion, the RV difference is approximately the same for both cases (flux correction on and off). In fact, when looking at the values, we noticed that they are exactly equal which means that the flux correction function was not able to correct anything at that level. Going further into investigation, we realized that this function is automatically switched off in ESPRESSO pipeline if the flux correction factor is outside the limits. The correction is only done if the flux factor is between 0.25 and 3. Looking to the headers of the output files of the pipeline, we realized that in the case where the ADC was at Max, the flux correction is indeed switched off since the minimum flux correction factor was lower than 0.25.  

\newpage
\section{Conclusions and future work}
In order to fully characterize the accuracy of the aforementioned atmospheric models, further observations are required. Since the tested spectral range is narrow, and since the dispersion has a very steep dependence on wavelength, it will be more interesting to use the same method and analysis for a broader spectral range that will go as low as possible to the blue. The accuracy of the measurements should be in the order of 1\% of the amount of dispersion. For example, for the ESPRESSO case, the dispersion between 380 nm and 780 nm is 2.4" for a zenith angle of 60$^{\circ}$; 1\% would correspond to a maximum error of 24 mas in the sky. In order to reach this accuracy, several measurements should be done. To do so, we submitted a proposal (P 103) to do some observations with UVES while putting the ADC into `simulation'/park position. We believe that we will be able to directly measure the differential refraction effects of the atmosphere down to the accuracy we need. \\
As for the impact of atmospheric dispersion on RV, it is clear from the analysis above that the difference on RV is proportional to the amount of residuals. It is important to understand the effect of small residuals (several tens of mas) to fully characterize the atmospheric dispersion effect. Since most high resolution astronomical instruments will be equipped with an ADC, and since most of the data reduction pipelines will take into consideration the flux correction, we are now trying to characterize the effects of residuals by generating synthetic spectra, affected by atmospheric dispersion residuals, that will be reduced in order to fully understand this effect. 

\acknowledgments % equivalent to \section*{ACKNOWLEDGMENTS}       
The first author is supported by an FCT fellowship (PD/BD/135225/2017), under the FCT PD Program PhD::SPACE (PD/00040/2012). The authors would also like to thank Gaspare Lo Curto for his help in acquiring data from ESPRESSO by seting the ADC at different positions.
% References
\bibliography{report} % bibliography data in report.bib
\bibliographystyle{spiebib} % makes bibtex use spiebib.bst

%\begin{thebibliography}{}
%\expandafter\ifx\csname natexlab\endcsname\relax\def\natexlab#1{#1}\fi
%\providecommand{\url}[1]{\href{#1}{#1}}
%\providecommand{\dodoi}[1]{doi:~\href{http://doi.org/#1}{\nolinkurl{#1}}}
%\providecommand{\doeprint}[1]{\href{http://ascl.net/#1}{\nolinkurl{http://%ascl.net/#1}}}
%\providecommand{\doarXiv}[1]{\href{https://arxiv.org/abs/#1}%{\nolinkurl{https://arxiv.org/abs/#1}}}

\end{document}